\documentclass[a4paper]{svmult}
\pdfoutput=1
\usepackage[hidelinks=true]{hyperref}
\usepackage{mathptmx}
\usepackage{latexsym}
\usepackage{tikz}
\usepackage{tgpagella}
\usepackage{geometry}
\usetikzlibrary{calc, shapes, matrix, arrows.meta, positioning}

\usepackage{listings}
\lstdefinestyle{customc}{
  belowcaptionskip=1\baselineskip,
  breaklines=false,
  language=Java,
  showstringspaces=false,
  fontadjust=true,
  flexiblecolumns=true,
  basicstyle=\small\ttfamily,
  numbers=left,
  numberstyle=\tiny, % the style that is used for the line-numbers
}
\usepackage{url}

\usepackage{xspace}
\newcommand{\vonda}{VOnDA\xspace}

\newcommand{\pal}{{\small\textsf{PAL}}\xspace}
\newcommand{\alize}{{\small\textsf{ALIZ-E}}\xspace}

\usepackage[textsize=small]{todonotes}
\usepackage[utf8]{inputenc}

\pgfdeclareimage[width=.99\columnwidth]{vondagui}{debugger}

\newcommand{\cmp}[2]{\begin{minipage}[h]{#1}\centering #2\end{minipage}}
\definecolor{code}{HTML}{FFE1B6}
\definecolor{midgray}{HTML}{B4B8AB}
\definecolor{darkblue}{HTML}{153243}
\definecolor{ivory}{HTML}{F4F9E9}
\definecolor{lightgray}{HTML}{EEF0EB}
\definecolor{meddarkblue}{HTML}{557C97}

\begin{document}
\title*{VOnDA: A Framework for Ontology-Based Dialogue Management}
\author{Bernd Kiefer and Anna Welker and Christophe Biwer}
\institute{Bernd Kiefer, Anna Welker, Christophe Biwer \at
German Research Center for Artificial Intelligence (DFKI), Saarbrücken, Germany
\email{kiefer@dfki.de, anna.welker@dfki.de, christophe.biwer@dfki.de}
}
\maketitle
\begin{tikzpicture}[remember picture,overlay]
  \path (current page.north) ++(0,-1)
  node[color=gray,anchor=north]{\cmp{1.1\textwidth}{\Large PRE-PRINT VERSION --
      presented at the Tenth International Workshop on Spoken Dialogue Systems
      Technology (IWSDS), April 24-26, 2019}};
\end{tikzpicture}
\newcommand{\abs}{%
We present \vonda, a framework to implement the dialogue management
functionality in dialogue systems. Although domain-independent, \vonda is
tailored towards dialogue systems with a focus on social communication, which
implies the need of a long-term memory and high user adaptivity. For these
systems, which are used in health environments or elderly care, margin of error
is very low and control over the dialogue process is of topmost importance.
%, which is a challenge for approaches based solely on machine-learning.
The same holds for commercial applications, where
customer trust is at risk. \vonda's specification and memory layer relies
upon (extended) RDF/OWL\footnote{Resource Description Framework \url{https://www.w3.org/RDF/}\\Web Ontology Language \url{https://www.w3.org/OWL/}}
, which provides a universal and uniform
representation, and facilitates interoperability with external data sources,
e.g., from physical sensors.}
\vspace*{-17ex}
\abstract*{\abs}
\abstract{\abs}
\vspace*{-2ex}
\section{Introduction}

Natural language dialogue systems are becoming more and more popular, be it as
virtual assistants such as Siri or Cortana, as Chatbots on websites providing
customer support, or as interface in human-robot interactions in areas ranging
from human-robot teams in industrial environments \cite{schwartz2016hybrid}
over social human-robot-interaction \cite{alize2010} to disaster response
\cite{kruijff2015tradr}.

A central component of most systems is the \emph{dialogue manager}, which
controls the (possibly multi-modal) reactions based on external triggers and
the current internal state. When building dialogue components for robotic
applications or in-car assistants, the system needs to take into account inputs
in various forms, first and foremost the user utterances, but also other sensor
input that may influence the dialogue, such as information from computer
vision, gaze detection, or even body and environment sensors for cognitive load
estimation.

In the following, we will describe \vonda, an open-source framework initially
developed to implement dialogue strategies for conversational robotic and
virtually embodied agents. The implementation mainly took place in the context
of the \alize and \pal projects, where a social robotic assistant supports
diabetic children managing their disease. This application domain dictates some
requirements that led to the decision to go for a rule-based system with
statistical selection and RDF/OWL underpinning.

Firstly, it requires a lot of control over the decision process, since mistakes
by the system are only tolerable in very specific situations, or not at all.
Secondly, it is vital to be able to maintain a relationship with the user over
a longer time period. This requires a long-term memory which can be efficiently
accessed by the dialogue system to exhibit familiarity with the user in various
forms, e.g., respecting personal preferences, but also making use of knowledge
about conversations or events that were part of interactions in past sessions.
For the same reason, the system needs high adaptability to the current user,
which means adding a significant number of variables to the state space. This
often poses a scalability problem for POMDP-based approaches, both in terms of
run-time performance, and of probability estimation, where marginal cases can
be dominated by the prominent situation. A third requirement for robotic
systems is the ability to process streaming sensor data, or at least use
aggregated high-level information from this data in the conversational system.

Furthermore, data collection for user groups in the health care domain is for
ethical reasons even more challenging than usual, and OWL reasoning offers a
very flexible way to access control.

\vonda therefore specifically targets the following design goals to support the
system requirements described before:

\begin{itemize}
  \addtolength{\itemsep}{-.6\itemsep}
\item Flexible and uniform specification of dialogue semantics, knowledge and
  data structures
\item Scalable, efficient, and easily accessible storage of interaction history
  and other data, like real-time sensor data, resulting in a large information
  state
\item Readable and compact rule specifications, facilitating access to the
  underlying RDF database, with the full power of a programming language
\item Transparent access to standard programming language constructs (Java
  classes) for simple integration with the host system
\end{itemize}

\vonda is not so much a complete dialogue management system as rather a
fundamental implementation layer for creating complex reactive systems, being
able to emulate almost all traditional rule- or automata-based frameworks. It
provides a strong and tight connection to a reasoning engine and storage, which
makes it possible to explore various research directions in the future.
% , from
% using streaming reasoning or probabilistic reasoning in connection with the
% rule system to adding a layer for developing agent functionality in the
% Belief-Desire-Intention paradigm.

In the next section, we review related
work that was done on dialogue
frameworks. In section \ref{sec:system}, we will give a high-level overview of
the \vonda framework, followed by a specification language synopsis. Section
\ref{sec:compiler} covers some aspects of the system
implementation. Section \ref{sec:applications} describes the application of the
framework in the \pal project's integrated system. The paper concludes with a
discussion of the work done, and further directions for research and
development.
%%% Local Variables:
%%% mode: latex
%%% TeX-master: "vonda"
%%% End:

\section{Related Work}

The existing frameworks to implement dialogue management
components roughly fall into two large groups, those that use symbolic
information or automata to specify the dialogue flow (IrisTK
\cite{skantze2012iristk}, RavenClaw \cite{bohus2009ravenclaw}, Visual SceneMaker
\cite{gebhard2012visual}), and those that mostly use statistical methods
(PyDial \cite{ultes2017pydial}, Alex \cite{jurvcivcek2014alex}). Somewhat in
between these is OpenDial \cite{lison2015developing}, which builds on
probabilistic rules and a Bayesian Network.

For reasons described in the introduction, \vonda currently makes only limited
use of statistical information. A meaningful comparison to purely learned
systems like PyDial or Alex therefore becomes more complex, and would have to
be done on an extrinsic basis, which we can not yet provide. We studied
comparable systems focusing mainly on two aspects: the specification of
behaviours, and the implementation of the dialogue memory / information state.

The dialogue behaviours in IrisTK and SceneMaker are specified using state
charts (hierarchical automata). Additional mechanisms (parallel execution,
history keeping, exception mechanisms like interruptive edges) make them more
flexible and powerful than basic state charts, but their flexibility and
generalisation capabilities are limited.

RavenClaw \cite{bohus2009ravenclaw} uses so-called \emph{task trees}, a variant
of flow charts that can be dynamically changed during run-time to implement
dialogue agents for different situations in the dialogue, and an \emph{agenda},
which selects the appropriate agent for the current dialogue state. The
resemblance to agent-based architectures using preconstructed plans is
striking, but the improved flexibility also comes at the cost of increased
complexity during implementation and debugging.

OpenDial \cite{lison2015developing} tries to combine the advantages of
hand-crafted systems with statistical selection, using probabilistic rules
which can be viewed as templates for probabilistic graphical models. The
parameters for the models can be estimated using previously collected data
(supervised learning), or during the interactions with reinforcement learning
techniques. Being able to specify structural knowledge for the statistical
selection reduces the estimation problem if only a small amount of data is
available, and allows to explicitly put restrictions on the selection process.

%\todo[inline]{How is vonda different or complementary to existing systems}
% Compare to OpenDIAL: big similarity, but more powerful going beyond
% condition/action, persistent memory, in fact, OpenDIAL can be emulated
% using \vonda

%Alex: check \url{https://github.com/UFAL-DSG/alex}
%IrisTK: statecharts
%- Long-term, multiple sessions
%- Well established

%%% Local Variables:
%%% mode: latex
%%% TeX-master: "vonda"
%%% End:

\section{High-Level System Description}
\label{sec:system}
\vonda follows the Information State / Update paradigm
\cite{traum2003information}. The information state represents everything the
dialogue agent knows about the current situation, possibly containing
information about dialogue history, the belief states of the participants,
situation data, etc., depending on the concrete system. Any change in the
information state will trigger a reasoning mechanism of some sort, which may
result in more changes in the information state, or outputs to the user or
other system components.

\vonda implements this paradigm by combining a rule-based approach with
statistical selection, although in a different way than OpenDial. The rule
specifications are close to if-then statements in programming languages, and
the information state is realised by an RDF store and reasoner with special
capabilities (HFC \cite{krieger2013efficient}), namely the possibility to
directly use $n$-tuples instead of triples. This allows to attach temporal
information to every data chunk \cite{Krieger:FOIS2012, krieger2014detailed}.
In this way, the RDF store can represent \emph{dynamic objects}, using either
\emph{transaction time} or \emph{valid time} attachments, and as a side effect
obtain a complete history of all changes.  HFC is very efficient in terms of
processing speed and memory footprint, and has recently been extended with
stream reasoning facilities. \vonda can use HFC either directly as a library,
or as a remote server, also allowing for more than one database instance, if
needed.
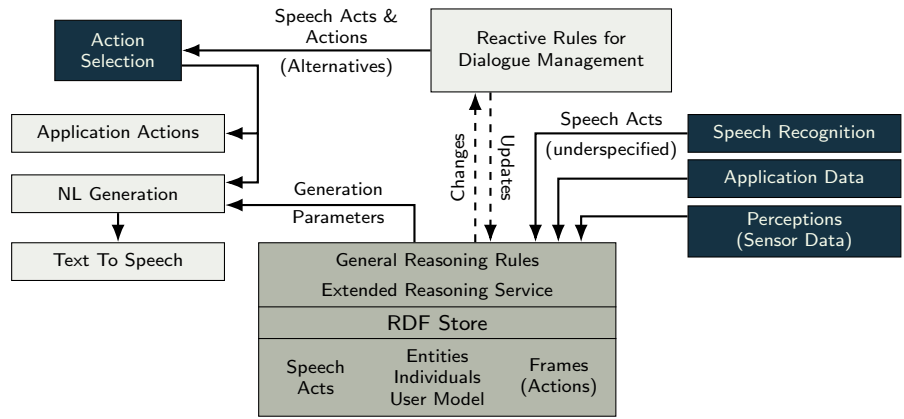
\begin{figure}[htb]
  \centering%\includegraphics[scale=0.33]{VondaArchitecture.png}\vspace*{-1ex}
  \centering
\scriptsize%
\begin{tikzpicture}[font=\sffamily,
  ampersand replacement=\&,
  box/.style={minimum width=2.8cm, minimum height=.5cm, thin, draw},
  gbox/.style={box, fill=lightgray},
  tbox/.style={box, fill=darkblue, text=ivory},
  ybox/.style={box, fill=lightgray},
  arr/.style={thick, -{Latex}},
  rbox/.style={box, fill=darkblue, text=ivory},
  ]

\path (5.7,3.8) node[gbox, minimum height=1.1cm](rr){\cmp{3cm}{Reactive Rules for\\Dialogue Management}};
\path (0,3.8) node[rbox, minimum width=1.6cm, minimum height=.8cm](as){\cmp{1.5cm}{Action\\Selection}};

\path (8.9,2.7) node[tbox](sr){Speech Recognition};
\path (8.9,2.1) node[tbox](ad){Application Data};
\path (8.9,1.4) node[tbox](per){\cmp{2.2cm}{Perceptions\\(Sensor Data)}};

\path (0,2.7) node[ybox](aa){Application Actions};
\path (0,1.9) node[ybox](nlg){NL Generation};
\path (0,1) node[ybox](tts){Text To Speech};

\path (4.2,0.1) node[box, fill=midgray, minimum width=4.7cm, minimum height=2.3cm](rdf){};
\path (4.2,0.19) node{\small\bf\sffamily RDF Store};
\path (4.2,1) node{General Reasoning Rules};
\path (4.2,0.6) node{Extended Reasoning Service};
\path (2.6,-.53) node{\cmp{.9cm}{Speech\\Acts}};
\path (4.2,-.53) node{\cmp{1.8cm}{Entities\\Individuals\\User Model}};
\path (5.8,-.53) node{\cmp{1.2cm}{Frames\\(Actions)}};

\path (rdf.east) ++(0,-.12) [draw, thin] coordinate(l1) -- (l1 -| rdf.west);
\path (rdf.east) ++(0,.3) [draw, thin] coordinate(l1) -- (l1 -| rdf.west);

\draw[arr] (rr) -- node[above,xshift=4mm,yshift=-4.5mm]{\cmp{2cm}{\scriptsize Speech Acts \& Actions\\[1ex]
    (Alternatives)}} (rr -| as.east);
\path (as.east) ++(0,-.2) [draw, arr] -- ++(1,0) coordinate(here) -|
(here |- aa) coordinate (there) |- (aa);
\path (nlg.east) ++ (0,.15) coordinate (nlg1);
\path (nlg.east) ++ (0,-.15) coordinate (nlg2);
\draw[arr] (there) |- (nlg1);
\path[arr, dashed] (rr.south) ++(-.8,0) coordinate(s1) [draw] -- node[rotate=-90,anchor=south]{\scriptsize Updates} (s1 |- rdf.north);
\path[arr, dashed] (rdf.north) ++(.5,0) coordinate(s2) [draw] -- node[rotate=90,anchor=south]{\scriptsize Changes} (s2 |- rr.south);
\path[arr] (rdf.north) ++(-.3,0) [draw] |- node[below, xshift=-1cm,yshift=4mm]{\cmp{2cm}{\scriptsize Generation\\[.8ex]Parameters}} (nlg2);

\draw[arr] (per.west) ++(0,0.2) -- ++(-1.4,0) coordinate(per1) -- (per1 |- rdf.north);
\draw[arr] (ad.west) -- ++(-1.7,0) coordinate(ad1) -- (ad1 |- rdf.north);
\draw[arr] (sr.west) -- node[above, yshift=-4.4mm]{\cmp{2cm}{\scriptsize Speech
    Acts\\[1ex](underspecified)}} ++(-2,0) coordinate(sr1) -- (sr1 |- rdf.north);

\draw[arr] (nlg) -- (tts);

\end{tikzpicture}
%%% Local Variables:
%%% mode: latex
%%% TeX-master: "vonda"
%%% End:
  \caption{\vonda Architecture}
\end{figure}
The initial motivation for using an RDF reasoner was our research interest in
multi-session, long-term interactions. In addition, this also allows processing
incoming facts in different layers. Firstly, there is the layer of custom
reasoning rules, which also comprises streaming reasoning, e.g., for real-time
sensor data, and secondly the reactive rule specifications, used mainly for
agent-like functionality that handles the behavioural part. This opens new
research directions, e.g., underpinning the rule conditions with a
probabilistic reasoner.

The RDF store contains the terminological and the dynamic knowledge:
specifications for the data types and their properties, as well as a hierarchy
of dialogue acts, semantic frames and their arguments, and the data objects,
which are instantiations of the data types. The data type specifications are
also used by the compiler to infer the types for property values (see section
\ref{sec:language}), and form a declarative API to connect new components,
e.g., for sensor or application data.

We are currently using the DIT++ dialogue act hierarchy \cite{bunt2012iso} and
shallow frame semantics along the lines of FrameNet
\cite{ruppenhofer2016framenet} to interface with the natural language
understanding and generation units. Our dialogue act object currently consist
of a dialogue act token, a frame and a list of key-value pairs as arguments
to the frame (\texttt{Offer(Transporting, what=tool, to=workbench)}).  While
this form of shallow semantics is enough for most applications, we already
experience its shortcomings when trying to handle, for example, social
talk. Since the underlying run-time core is already working with full-fledged
feature matrices, only a small syntax extension will be needed to allow for
nested structures.

A set of \emph{reactive condition-action rules} (see figure~\ref{fig:code}) is
executed whenever there is a change in the information state. These changes are
caused by incoming sensor or application data, intents from the speech
recognition, or expired timers.  Rules are labelled if-then-else statements,
with complex conditions and shortcut logic, as in Java or C. The compiler
analyses the base terms and stores their values during processing for dynamic
logging. A rule can have direct effects, like changing the information state or
executing system calls. Furthermore, it can generate so-called
\emph{proposals}, which are (labelled) blocks of code in a frozen state that
will not be immediately executed, similar to closures.

All rules are repeatedly applied until a fixed point is reached where no new
proposals are generated and there is no information state change in the last
iteration. Subsequently, the set of proposals is evaluated by a statistical
component, which will select the best alternative. This component can be
exchanged to make it as simple or elaborate as necessary, taking into account
arbitrary features from the data storage.

\begin{figure}[htb]
  \centering\tiny%
\begin{tikzpicture}[font=\bf\sffamily,
  box/.style={minimum width=4.5cm, minimum height=.35cm, draw=gray, text=black},
  gbox/.style={box, fill=lightgray},
  arr/.style={thick, -{Stealth}},
  larr/.style={thick, {Stealth}-},
  oarr/.style={thick, -{Stealth}},
  bbox/.style={box, fill=meddarkblue, text=ivory, draw},
  rbox/.style={node distance=0.3cm, font=\ttfamily},
  fbox/.style={node distance=4.2cm, font=\ttfamily},
  rlabel/.style={right, xshift=1mm, font=\sl},
  llabel/.style={left, xshift=-1mm, font=\sl}
]
\node[gbox](rmn) at (0,0.2) { Rule Module Class N };
\node[rbox, right= of rmn](rmnj){RuleModuleN.java};
\draw[larr] (rmnj) --  ++(1.5,0);
\node[fbox, right= of rmn](rmnr){RuleModuleN.rudi};
\draw[oarr] (rmnr) --  ++(-1.5,0);

\node[minimum width=4.5cm](ddd) at (0,-.2) { \normalsize ... };
\node[rbox, right= of ddd]{\small ...};
\node[fbox, right= of ddd]{\small ...};

\path (rmn) ++(0,-.75) node[gbox](rm1) { Rule Module Class 1 };
\node[rbox, right= of rm1](rm1j){RuleModule1.java};
\draw[larr] (rm1j) --  ++(1.5,0);
\node[fbox, right= of rm1](rm1r){RuleModule1.rudi};
\draw[oarr] (rm1r) --  ++(-1.5,0);

\path (rm1) ++(0,-.75) node[gbox](tlrc) { Top Level Rule Class };
\node[rbox, right= of tlrc](maj){MyAgent.java};
\draw[larr] (maj.east) ++(.35,0) --  ++(.72,0);
\node[fbox, right= of tlrc](mar){MyAgent.rudi};
\draw[oarr] (mar.west) -- ++(-.7,0);
\draw[arr] (tlrc.north) ++ (1,0) coordinate (tn) -- node[rlabel]{calls} (rm1.south -| tn);
\draw[arr] (rm1.south) ++ (-1,0) coordinate (rs) -- node[llabel]{imports} (tlrc.north -| rs);

\path (tlrc) ++(0,-.75) node[box, fill=ivory](caai) {Concrete Agent API Implementation};
\node[rbox, right= of caai]{MyAgentBase.java};
\draw[arr] (caai) -- node[rlabel]{extends} (tlrc);

\path (caai) ++(0,-.75) node[box, fill=darkblue, text=ivory](caid) {Common Agent API Interface Description}
 ++ (0,-.35) node[bbox](caii) {Common Agent API Implementation}
% ++ (0,-.35) node[bbox] {VOnDA Runtime}
 ++ (0,-.45) node[bbox, minimum height=.6cm](cb) {
\begin{minipage}{2cm}\centering Main event\\processing loop\end{minipage}
\begin{minipage}{2cm}\centering RDF Object Access\end{minipage}
}
 ++ (0, -.4) node[bbox] {Dialogue Act Creaton / Comparison}
 ++ (0,-0.35) node[bbox](nlp){
\begin{minipage}{2cm}\centering NLG\end{minipage}
\begin{minipage}{2cm}\centering NLU\end{minipage}
};
\node[rotate=90,yshift=2mm] at (cb.west){\scriptsize VOnDA Runtime};

\node[rbox, right= of caii]{Agent.java};
\node[rbox, right= of caid]{Agent.rudi};
\draw[arr] (caid) -- node[rlabel]{extends} (caai);
\draw (cb) ++(0,.3) -- +(0,-.51);
\draw (nlp) ++(0,.18) -- +(0,-.36);

\path (nlp) ++(0,-.75)
+(-1.125,0) node[fill=lightgray, minimum width=2.25cm, minimum height=.35cm]{}
+(1.125,0) node[fill=darkblue, text=ivory, minimum width=2.25cm, minimum height=.35cm]{}
+(0,0) node[box](ci) {\hspace{2.7ex}Client\color{ivory}\ \ Interface };
\node[rbox, right= of ci]{StubClient.java / MyClient.java};

\draw[arr] (ci.north) ++ (1,0) coordinate (ci1) -- (nlp.south -| ci1);
\draw[arr] (nlp.south) ++ (-1,0) coordinate (nl1) -- (ci.north -| nl1);

\node[minimum width=2cm, minimum height=.7cm, %draw,
      fill=meddarkblue, text=ivory, rotate=90](vc) at (5.3,-.6) {\bf\sffamily VOnDA Compiler};

\node[minimum width=1.3cm, minimum height=.8cm, %draw,
      fill=meddarkblue, text=ivory](dbg) at (7.3,-4.9)
      {\begin{minipage}{1.3cm}\centering\bf\sffamily VOnDA\\Debugger\end{minipage}};

% SUPER TEMPLATE FÜR DIE "TONNE"
\path [fill=midgray, text=black, draw=black] (5.3, -4)
   node[minimum height=1.24cm,minimum width=1.5cm] (db) {\cmp{1.3cm}{RDF\\Database}}
   ++(-.75,0.5)
   -- ++(0,-.9)
   arc [start angle=180, delta angle=180, x radius=.75cm, y radius=1.2mm]
   -- ++(0,.9)
   arc[start angle=0, delta angle=360, x radius=.75cm, y radius=1.2mm];

\draw[arr] (vc) --
  node[above, rotate=90, xshift=1.1mm]{Class \& Predicate}
  node[below, rotate=90, xshift=1.1mm]{Definitions} (db);
\draw[arr, {Stealth}-{Stealth}] (db.west)
       -- node[above]{Belief State Data} (db.west -| cb.east);

\draw[arr](dbg) -- (nlp);
\draw[arr](dbg) -- (db);
\draw[arr](dbg) -- (vc.south west);
\end{tikzpicture}

%%% Local Variables:
%%% mode: latex
%%% TeX-master: "vonda"
%%% End:
  \caption{\label{fig:agent}A schematic \vonda agent}
\end{figure}
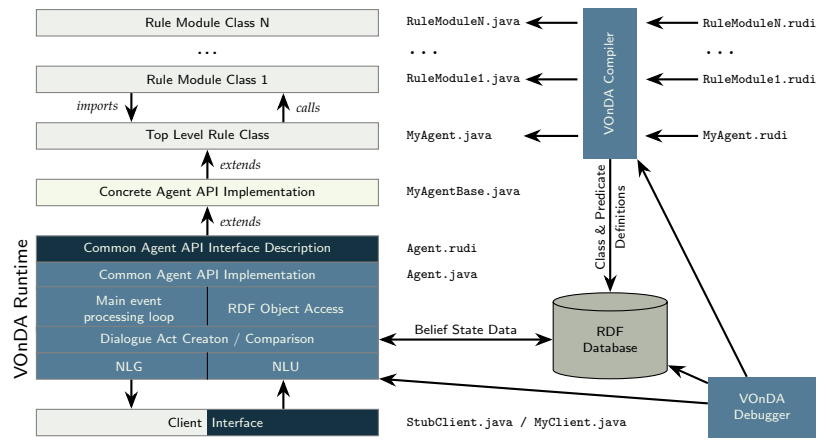
A \vonda project consists of an ontology, a custom extension of the abstract
\texttt{Agent} class (the so-called \emph{wrapper class}), a client interface
to connect the communication channels of the application to the agent, and a
set of rule files that are arranged in a tree, using \texttt{import}
statements. The blue core in Figure~\ref{fig:agent} is the run-time system
which is part of the \vonda framework, while all elements above are application
specific parts of the agent. A \texttt{Yaml} project file contains all
necessary information for compilation: the ontology, the wrapper class, the
top-level rule file and other parameters, such as custom compile commands.

The ontology contains the definitions of dialogue acts, semantic frames, class
and property specifications for the data objects of the application, and other
assertional knowledge, such as specifications for ``forgetting'', which could
be modeled in an orthogonal class hierarchy and supported by custom deletion
rules in the reasoner.

Every rule file can define variables and functions in \vonda syntax which are
then available to all imported files. The methods from the wrapper class are
available to all rule files.

The current structure assumes that most of the Java functionality that is used
inside the rule files will be provided by the \texttt{Agent} superclass. There
are, however, alternative ways to use other Java classes directly, with support
for the same type inference as for RDF classes.

%%% Local Variables:
%%% mode: latex
%%% TeX-master: "vonda"
%%% End:

%\section{Implementation}
\section{Dialogue Specification Language}
\label{sec:language}

\vonda's rule language at first sight looks very similar to Java/C++. However,
there are a number of specific features which make it convenient for the
implementation of dialogue strategies. Maybe the most important one is the
handling of RDF objects and classes, which can be treated similarly to those of
object oriented programming languages, including the (multiple) inheritance and
type inference that are provided by the RDF class hierarchies.
\vspace*{-2ex}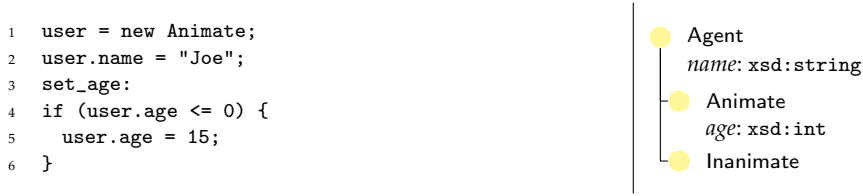
\begin{figure}[htb]
\hspace*{4ex}
\begin{minipage}{0.5\columnwidth}
\small\begin{lstlisting}
user = new Animate;
user.name = "Joe";
set_age:
if (user.age <= 0) {
  user.age = 15;
}
\end{lstlisting}\end{minipage}\ \vrule\hspace{1ex}
\begin{minipage}{0.44\columnwidth}
    \small\begin{tikzpicture}[
  blob/.style={circle, fill=yellow!50!white, minimum width=2mm},
  txt/.style={node distance=2.5mm,font=\sffamily}]
  \draw (0,0) node (agent) [blob]{};
  % BEWARE: RIGHT OF= IS DEPRECATED, DON'T USE IT
  \node (agtxt) [right= 0.1 of agent, txt] {Agent};
  \node (name) [below= 0.4 of agtxt.west, anchor=west]{\emph{name}: \texttt{xsd:string}};
  \node (animate) [blob, below=0.7 of agtxt.west]{};
  \node (antxt) [right= 0.1 of animate, txt] {Animate};
  \node (name) [below= 0.4 of antxt.west, anchor=west]{\emph{age}: \texttt{xsd:int}};
  \node (inanimate) [blob, below= 0.5 of animate, node distance=9mm]{};
  \node (intxt) [right= 0.1 of inanimate, txt] {Inanimate};
  \draw (agent) |- (animate);
  \draw (agent) |- (inanimate);
\end{tikzpicture}
\end{minipage}
  \caption{Ontology and \vonda code}
  \label{fig:rdfobjects}
\end{figure}\vspace*{-2ex}

Figure \ref{fig:rdfobjects} contains an example of \vonda code, and how it
relates to RDF type and property specifications, schematically drawn on the
right. The domain and range definitions of properties are picked up by the
compiler and used in various places, e.g., to infer types, do automatic code or
data conversions, or create ``intelligent'' boolean tests, such as the one in
line 4, which will expand into two tests, one testing for the existence of the
property for the object, and in case that succeeds, a test if the value is
greater than zero. If there is a chain of more than one field resp. property
access, every part is tested for existence in the target code, keeping the
source code as concise as possible. Also, for reasons of brevity, the type of a
new variable needs not be given if it can be inferred from the value assigned
to it.

New RDF objects can be created with \texttt{new}, similar to Java objects; they
are immediately reflected in the database, as are all changes to already
existing objects.

Many operators are overloaded, especially boolean operators such as
\textbf{\texttt{<=}}, which compares numeric values, but can also be used to test if an
object is of a specific class, for subclass tests between two classes, and for
subsumption of dialogue acts.
%\vspace*{-2ex}

\begin{figure}[htb]
\centering\begin{minipage}{.91\textwidth}\begin{lstlisting}
if (!saidInSession(#Greeting(Meeting)) {
  timeout("wait_for_greeting", 7000){ //Wait 7 secs before taking initiative
    if (! receivedInSession(#Greeting(Meeting))
      propose("greet") {
        da = #InitialGreeting(Meeting);
        if (user.name) da.name = user.name;
        emitDA(da);
      }
  }

  if (receivedInSession(#Greeting(Meeting))
    propose("greet_back") { // We assume we know the name by now
      emitDA(#ReturnGreeting(Meeting, name={user.name}));
    }
}
\end{lstlisting}\end{minipage}\vspace*{-1ex}
  \caption{\vonda code example}
  \label{fig:code}
\end{figure}%\vspace*{-2ex}
There are two statements with a special syntax and semantics: \texttt{propose}
and \texttt{timeout}. \texttt{propose} is \vonda's current way of implementing
probabilistic selection. All (unique) propose blocks that are in active rule
actions are collected, frozen in the execution state in which they were
encountered, such as closures known from functional programming languages. When
rule processing stops, a statistical component picks the ``best'' proposal and
its closure is executed.

\texttt{timeout}s also generate closures, but with a different purpose. They
can be used to trigger proactive behaviour, or to check the state of the system
after some time period, or in regular intervals. A timeout will only be created
if there is no active timeout with that name.

Figure~\ref{fig:code} also contains an example of the short-hand notation
for shallow semantic structures (starting with \textbf{\texttt{\#}}).  Since
they predominantly contain constant (string) literals, this is the default when
specifying such structures. The special syntax in
\texttt{user=\{user.name\}} allows to insert the value of expressions into the
literal, similar to an \emph{eval}.

This section only described the most important features of \vonda's syntax. For
a detailed description, the reader is referred to the user documentation\footnote{\url{https://github.com/bkiefer/vonda/blob/master/doc/master.pdf}}.

%%% Local Variables:
%%% mode: latex
%%% TeX-master: "vonda"
%%% End:

\section{Compiler / Run-Time Library}
\label{sec:compiler}

The compiler turns the \vonda source code into Java source code using the
information in the ontology. Every source file becomes a Java class. Although
the generated code is not primarily for the human reader, a lot of care has
been taken in making it still understandable and debuggable. The compile
process is separated into three stages: parsing and abstract syntax tree
building, type checking and inference, and code generation.

The \vonda compiler's internal knowledge about the program structure and the
RDF hierarchy takes care of transforming the RDF field accesses into reads from
and writes to the database. Beyond that, the type system, resolving the exact
Java, RDF or RDF collection type of (arbitrary long) field accesses, automatically
performs the necessary casts for the ontology accesses.

%\section{Run-Time Library}

The run-time library contains the basic functionality for handling the rule
processing, including the proposals and timeouts, and for the on-line
inspection of the rule evaluation. There is, however, no blueprint for the main
event loop, since that depends heavily on the host application. It also
contains methods for the creation and modification of shallow semantic
structures, and especially for searching the interaction history for specific
utterances. Most of this functionality is available through the abstract
\texttt{Agent} class, which has to be extended to a concrete class for each
application.

There is functionality to directly communicate with the HFC database using
queries, in case the object view is not sufficient or too awkward. The natural
language understanding and generation components can be exchanged by
implementing existing interfaces, and the statistical component is connected by
a message exchange protocol. A basic natural language generation engine based
on a graph rewriting module is already integrated, and is used in our current
system as a template based generator. The example application also contains a
VoiceXML based interpretation module.

%%% Local Variables:
%%% mode: latex
%%% TeX-master: "vonda"
%%% End:

\subsubsection*{Debugger / GUI}
\label{sec:debugger}
\vonda comes with a GUI \cite{rudibuggerThesis} that helps navigating, compiling and editing the source
files belonging to a project. It uses the project file to collect all the
necessary information.

\begin{figure}[thb]
  \pgfuseimage{vondagui}
  \caption{The \vonda GUI window}\vspace*{-1ex}
\end{figure}

% Writing / Compilation part

Upon opening a project, the GUI displays the project directory (in a
\textit{file view}).  The user can edit rule files from within the GUI or with
an external editor like Emacs, Vim, etc.  and can start the compilation
process. After successful compilation, the project view shows what files are
currently used, and marks the top-level and the wrapper class files. A second
tree view (\emph{rule view}) shows the rule structure in addition to the module
structure. Modules in which errors or warnings were reported during compilation
are highlighted, and the user can quickly navigate to them using context menus.

% Live system part

Additionally, the GUI can be used to track what is happening in a running
system. The connection is established using a socket to allow remote debugging.
In the rule view, multi-state check boxes are used to define which rules should
be observed under which conditions. A rule can be set to be logged under any
circumstances, not at all or if its condition evaluated to true or to
false. Since the rules are represented in a tree-like structure, the logging
condition can also be set for an entire subgroup of rules, or for a whole
module. The current rule logging configuration can be saved for later use.

The \emph{logging view} displays incoming logging information as a sortable
table. A table entry contains a time stamp, the rule's label and its
condition. The rule's label is coloured according to the final result of the
whole boolean expression. Each base term of the condition is coloured
accordingly, or greyed out if short-cut logic led to premature failure or
success of the expression. Inspecting the live system helps pin-point problems
when the behaviour is not as expected. The log shows how the currently active
part of the information state is processed, and the window offers easy
navigation using the mouse from the rule condition to the corresponding source
code.

%%% Local Variables:
%%% mode: latex
%%% TeX-master: "vonda"
%%% End:

\section{Applications}
\label{sec:applications}
\vonda is  used in the integrated system of the EU project \pal
\cite{palWebsite}, which uses human-robot interaction to support children with
diabetes type 1 in coping with their disease. Children interact with a real NAO
robot\footnote{Softbank Robotics~\scriptsize{\url{https://www.ald.softbankrobotics.com}}}, or with an Android app that connects to the
core system and exhibits a virtual character that is as similar to the robot as
possible, also in its behaviour.

The dialogue component, which is largely responsible for the agent's behaviour,
is implemented using the \vonda framework. In addition, HFC, the RDF store that
\vonda builds upon, is the main database of the system, storing all relevant
information and being the central data exchange hub. The system runs as a
cloud-based robotic solution, spawning a new system instance for every user. It
has been successfully tested with more than 40 users at a time on a medium
sized virtual machine\footnote{4 core Xeon E5-2683@2.00GHz, 16 GB RAM} with only
moderate load factors, giving a positive indication of the scalability of HFC
and the \vonda approach.

There are two helper modules integrated into the dialogue component which quite
extensively exploit the connection between the database and the rule part,
namely the \emph{Episodic Memory} and the \emph{Targeted Feedback}. While the
targeted feedback reacts to current events in the running session, like
entering a bad or good glucose value, or the current achievement of a task, the
episodic memory aggregates data from the past and eventually converts them into
so-called episodes that are used for interactions in subsequent sessions. Both
are only triggered if relevant changes in the database occur, for example
incoming data from the MyPAL app about games or achievements, and serve
different conversational purposes, namely showing familiarity with the user and
her/his everyday life, versus reacting to current positive or negative
incidents.

\vonda has also been used in a recent project aiming to implement a
generalised, ontology-based approach to open-domain talk \cite{welker2017}. The
Smoto system uses an additional HFC server running WordNet
\cite{miller1995wordnet, fellbaum1998wordnet} as semantic database, thereby
gaining knowledge about semantic concepts that can be used in the dialogue and
to find appropriate reactions on arbitrary user input.

\section{Discussion and Further Work}

We believe that there are still many interesting application areas for hybrid
statistical and hand crafted systems, e.g., if they are relatively small, or
there is little domain-specific data available. Many currently deployed systems
that build on much simpler technology like VoiceXML can certainly profit from
hybrid approaches such as OpenDial or \vonda.

% As \cite{stoyanchev2016rapid} puts it: \emph{``Indeed, the bulk of currently
% deployed dialogue systems continue to rely on traditional hand-crafted
% finite-state or rule-based approaches to dialogue management using commercial
% or proprietary tools generating VoiceXML.''}.

\vonda is under active development. We designed it such that it can be
integrated in most applications and opens many ways for improvements and
additions. As a rule-based framework that is close to being a programming
language, \vonda is able to completely emulate the automata-based frameworks. In fact,
we are currently working on a graphical editor à la SceneMaker and the
precompilation of hierarchical state charts into \vonda code.  We hope this
will facilitate the implementation of new applications for inexperienced users
and help with rapid prototyping, while retaining the greater flexibility and
modularization capabilities. In this way, we combine the intuitive way of
specifying simple strategies with the full flexibility of the framework.

\vonda could also be used to implement modules that simulate the agents of
RavenClaw. To get a functionality similar to RavenClaw's agenda, its
action selection module would have to be implemented as a dialogue state
tracker, activating the most probable agent at each dialogue step.

Using the well-established RDF/OWL standard as specification layer makes it
very easy to add or change application specific data structures, especially
because of the existing tool support. We already use the reasoning facilities
for type and partially for temporal inference, but given the possibility of
attaching also confidence or credibility information to the RDF data, a more
integrated probabilistic approach with soft preconditions could be implemented,
e.g., on the basis of Dempster-Shafer theory
\cite{dempster2008}. Moreover, additional meta knowledge, such as
trustworthiness or validity periods could be declared using multiple
inheritance, which opens many interesting research directions.

Other next steps will be the addition of default adaptors for obviously needed
external modules like automatic speech recognition, more flexible language
understanding, and the like. We will also work on the improvement of the GUI,
including features such as a watch window and/or a timeline to track changes of
specific values in the database, and a tool that analyses the dependencies
between rules on the basis of the conditions' base terms.

From the research perspective, there are two very interesting lanes:
integrating probabilistic reasoning as a first-class option, which is directly
integrated with the rule conditions, and adding an additional layer to
facilitate the implementation of BDI-like agents, to study the connections and
dependencies between conversational and non-conversational behaviours.
\vspace*{-1ex}\subsubsection*{Source Code and Documentation}
The \vonda core system can be downloaded at
{\small\url{git@github.com:bkiefer/vonda.git}}. The main page has detailed
instructions for the installation of external dependencies. The debugger
currently lives in a separate project:
{\small\url{git@github.com:yoshegg/rudibugger.git}}. Both projects are licensed
under the Creative Commons Attribution-NonCommercial 4.0 International
License\footnote{\url{http://creativecommons.org/licenses/by-nc/4.0/}}, and are
free for all non-commercial use. A screen cast showing the GUI functionality
and the running PAL system is available at
{\small\url{https://youtu.be/nSotEVZUEyw}}.
\vspace*{-1ex}\subsubsection*{Acknowledgements}
The research described in this paper has been funded by the
Horizon 2020 Framework Programme of the European Union
within the project \textsf{PAL} (Personal Assistant for healthy Lifestyle)
under Grant agreement no. 643783.
\vspace*{1ex}

\noindent
This paper is dedicated to our colleague and friend Hans-Ulrich Krieger, the
creator of HFC. Hope you found peace, wherever you are.
\bibliography{vonda}
\bibliographystyle{spmpsci}

\end{document}